# Arctic amplification metrics

Richard Davy[1*], Linling Chen[1] and Edward Hanna[2]

[1]*Nansen Environmental and Remote Sensing Centre, Thormøhlensgate 47, 5006 Bergen, Norway*

*[*Richard.davy@nersc.no](*Richard.davy@nersc.no)*

[2]*School of Geography, University of Lincoln, UK*

**Abstract.** One of the defining features of both recent and historical cases of global climate change is Arctic Amplification (AA). This is the more rapid change in the surface air temperature (SAT) in the Arctic compared to some wider reference region, such as the Northern Hemisphere (NH) mean. Many different metrics have been developed to quantify the degree of AA based on SAT anomalies, trends and variability. The use of different metrics, as well as the choice of dataset to use can affect conclusions about the magnitude and temporal variability of AA. Here we review the established metrics of AA to see how well they agree upon the temporal signature of AA, such as the multi-decadal variability, and assess the consistency in these metrics across different commonly-used datasets which cover both the early and late 20$^{th}$ century warming in the Arctic. We find the NOAA 20$^{th}$ Century Reanalysis most closely matches the observations when using metrics based upon SAT trends ($A_2$), variability ($A_3$) and regression ($A_4$) of the SAT anomalies, and the ERA 20$^{th}$ Century Reanalysis is closest to the observations in the SAT anomalies ($A_1$) and variability of SAT anomalies ($A_3$). However, there are large seasonal differences in the consistency between datasets. Moreover, the largest differences between the century-long reanalysis products and observations are during the early warming period, likely due to the sparseness of the observations in the Arctic at that time. In the modern warming period, the high density of observations strongly constrains all the reanalysis products, whether they include satellite observations or only surface observations. Thus, all the reanalysis and observation products produce very similar magnitudes and temporal variability in the degree of AA during the recent warming period.

## 1. Introduction

Metrics of Arctic Amplification (AA) allow us to distinguish the periods in which there is the greatest difference between the surface air temperature (SAT) response to climate change in the Arctic and in the Northern Hemisphere (NH) as a whole. There have been two periods in the 20$^{th}$ century which have been identified as exhibiting AA: in the 1920s to 1940s (Yamanouchi, 2010; Wood and Overland, 2010), and at the end of the 20$^{th}$ century continuing into the 21$^{st}$ century (Overland et al., 2008; Serreze and Barry, 2011, Overland et al. 2016a). There are also numerous examples of AA in paleoclimate records and simulations (Masson-Delmotte et al., 2006; Miller et al., 2010; Brigham-Grette et al., 2013). In addition to this multi-decadal and longer-period variability in the observed AA, we can also see strong seasonal differences in the magnitude of AA (Serreze et al., 2009; Serreze and Barry, 2011). AA is generally strongest in winter (Screen and Simmonds, 2010; Serreze and Barry, 2011), as seen for example by much stronger Greenland warming in winter since the early 1990s (Hanna et al. 2012), when the SAT is more sensitive to changes in thermal forcing (Davy and Esau, 2016). Such changes in thermal forcing have been ascribed to feedback effects due to a reduction in sea-ice, the Planck feedback, changes in atmospheric water vapour, cloud cover, or increased advection (Serreze and Francis, 2006; Serreze and Barry, 2011; Screen et al., 2012). However, a recent analysis of global climate models has shown that much of the AA in the surface air temperature is due to local temperature feedbacks (Pithan and Mauritsen, 2014) whereby the persistent stable stratification in the atmospheric boundary layer traps excess heat in a shallow layer of air, leading to enhanced warming compared to the global average (Lesins et al., 2012; Esau et al., 2012).

There are several metrics which can be used to describe the difference in temperature response that is the signature of AA. It is necessary to have different metrics given that these metrics are often used to select periods of interest to study different climate processes. For example, those studies which concentrate on positive climate feedbacks involved in AA will want to focus on periods of rapid Arctic warming, and so may use a metric for AA based upon the rate of warming in the Arctic. Meanwhile those studies looking at the effect on the general circulation of a warmer Arctic may choose to use a metric based on the surface air temperature difference between the Arctic and the NH.

Arguably the simplest metric, and the most commonly used, is defined as the difference in the SAT anomalies in two regions. SAT anomalies are calculated by removing the climatology from the temperature time series at each location around the world and then a metric for AA is created by taking an Arctic-average temperature anomaly and comparing it to some reference temperature anomaly (e.g. the NH mean). This comparison may be taken to be the difference or the ratio of the two anomalies (Crook et al., 2011; Kobashi et al., 2013; Francis and Vavrus, 2015). However, there is a danger when using a ratio of two variables as a metric of AA when the denominator can approach zero, as can sometimes happen where anomalies are used (Hind et al., 2016). Metrics based on temperature anomalies are subject to a high degree of temporal variability at monthly timescales and longer due to the relatively large natural variability of the SAT in the Arctic (Legate and Willmot, 1990; Jones and Moberg, 2003).

An alternative metric for AA that was recently proposed uses the ratio of the absolute values of 30-year linear trends in the SAT over the Arctic and the NH (Johannessen et al., 2016). This metric has the advantage that, owing to the thirty-year running-window used to calculate the trends, it does not have the high temporal variability that is found in the metrics based on temperature anomalies, so it can readily be used to assess the behaviour of the Arctic on multi-decadal and longer timescales. This is an appealing metric for application to many climate studies which wish to focus on the climate-processes during extended periods of rapid Arctic warming; however, there are a couple of challenges with using this metric. Firstly, linear trends can be sensitive to outlier data points, especially when these are close to the beginning or end of the record, and so such outliers can potentially introduce a high degree of variability to the metric (Liebmann et al., 2010). Secondly, in taking the ratio of two trends we must account for the uncertainty in both of the linear regressions when determining the value of the AA metric. This can lead to large uncertainties in the magnitude of the metric in addition to missing values at times when we do not establish a statistically-significant difference between the two trends, which can make it harder to assess temporal variability.

One way around this issue is to use the inter-annual variability of the SAT to define the degree of AA. Note that here we use the term variability to refer to the standard deviation of anomalies regardless of their temporal structure (Suteanu, 2015). There is a larger inter-annual variability in the SAT in the Arctic than in the globe as a whole (Legate and Willmot, 1990; Jones and Moberg, 2003). This is partly due to synoptic activity and radiative-feedbacks in the Arctic (Stone, 1997; Vihma 2014), the effects of which are amplified by the persistent stable stratification found in the high latitudes, leading to a higher sensitivity of the SAT (Esau et al., 2012; Pithan and Mauritsen, 2014). So the difference or ratio of the SAT variability in the Arctic and some reference region can also be used as a metric of AA (Kobashi et al., 2013). This has the advantage that it is a temporally-continuous metric which allows us to fully assess seasonal, inter-annual and multi-decadal temporal variability of AA. One can also use either differences or ratios to describe the AA because the reference variability will not approach zero and so make the ratio unconstrained.

The last metric that we assess here is the coefficient of linear regression of Arctic SAT anomalies against NH SAT anomalies (Bekryaev et al., 2010). This metric has the distinct advantage of being more stable over the choice of period as compared to other metrics of AA. This is because it reduces the influence of variability, and especially multi-decadal variability, on the signal of AA. Therefore it is a relatively robust metric for AA across a range of timescales.

In this paper we do not seek to address the causes of AA, but instead we review a set of established metrics for AA and assess how consistent they are across a range of existing datasets and time periods. We also assess the sensitivity of the different metrics to choices of the period of analysis and the choice of dataset. In Section 2 we present our methodologies and datasets; in Section 3 we present the results from the different metrics; and in Section 4 we discuss some important considerations when choosing a metric to assess AA.

## 2. Methods

Here we present the results from four different datasets: two observational records and six reanalysis. The two observational gridded temperature datasets are the NASA Goddard Institute for Space Studies' "GISTEMP" (Schmidt et al., 2016) and the globally-extended version of the Met Office Hadley centre's HadCRUT4 temperature dataset by Cowtan and Way (2014), Had4krig_v2. The six reanalysis datasets we used were the European Centre for Medium-Range Weather Forecasts (ECMWF) 20$^{th}$ Century atmosphere-only reanalysis, ERA20C (Poli et al., 2016); the ECMWF's interim reanalysis, ERAint (Dee et al., 2011); the Japanese Meteorological Agency's JRA55 (Kobayashi, et al., 2015); NASA's Global Modeling and Assimilation Office's MERRA2 (Gelaro et al., 2017); the National Centre for Atmospheric Research's CFSR (Saha et al., 2014); and the National Oceanic and Atmospheric Administration's 20$^{th}$ Century reanalysis version 2C, 20CRv2c (Compo et al., 2011). The observational datasets have a monthly resolution, while the reanalysis datasets are available at daily resolution. The reanalysis products greatly vary in which observations are assimilated. The ERAint, MERRA2, JRA55, and CFSR analysis include satellite observations in the analyses, whereas the 20CRv2c assimilated surface pressure, monthly sea surface temperature and sea ice observations and the ERA20C used surface pressure and marine wind observations.

The inter-comparison of these two century-long reanalyses with the shorter-period reanalysis allows us to compare the effect of including more than surface observations on the representation of AA. Although it should be emphasized that surface observations of pressure and marine winds in the Arctic are sparse for the early warming period with no coverage over the ocean, so we might reasonably expect the models to deviate in the representation of Arctic climate prior to the 1950s. When observational data are sparse the climate in the reanalysis product becomes strongly dependent on the underlying dynamical model and consequentially will adopt any biases innate to the dynamical model. This severely limits are interpretation of the early warming period as there were very few observations from the Arctic Ocean at this time (Polyakov et al., 2003; Delworth and Knutson, 2000).

For each of these datasets we calculate four metrics of AA, henceforth labelled $A_1$, $A_2$, $A_3$ and $A_4$ and these are summarised in Table 1. The first metric is the difference in SAT anomalies in the Arctic and the NH as a whole ($A_1$); the second is the ratio of the magnitude of the 21-year linear temperature trend in the Arctic to that in the NH ($A_2$); the third is the ratio in the inter-annual SAT variability between the Arctic and the NH as a whole ($A_3$); and the fourth is the slope of the linear regression between the Arctic and NH SAT anomalies ($A_4$).

SAT anomalies were calculated by subtracting the common reference period 1981-2010 climatological monthly means from the full time series for the respective calendar months. Annual anomalies were calculated by taking the mean of the monthly anomalies. Seasonal anomalies were calculated by taking the means of three monthly anomalies using standard definitions of winter (December, January and February) and summer (June, July and August). The Arctic and NH SAT anomalies were then calculated by applying area-weighted averages over the respective regions. These two time series, Arctic and NH SAT anomalies, were then used to calculate the different AA metrics. The 21-year linear trends were calculated by fitting a linear regression in time to the SAT anomalies and the regressions were tested for significance using a two-tailed student-t distribution. The interannual variability was calculated from the standard deviation of the SAT anomalies. The coefficient of linear regression between the two SAT anomaly time-series in $A_4$ was calculated using a least-squares linear regression on the monthly anomalies in a 21-year window and for each regression result statistical significance was tested as above.

| Metric ID | Definition | Reference |
|---|---|---|
| $A_1$ | {SAT anomaly in Arctic} – {SAT anomaly in NH} | Francis and Vavrus, 2015 |
| $A_2$ | \|SAT 21-year linear trend in Arctic\| / \|SAT 21-year linear trend in NH\| | Johannessen et al., 2016 |
| $A_3$ | {Inter-annual SAT variability in Arctic} / { Inter-annual SAT variability in NH} | Kobashi et al., 2013 |
| $A_4$ | Coefficient of linear regression between Arctic and NH SAT anomalies | Bekryaev et al., 2010 |

**Table 1**. Summary of metrics for AA used in this manuscript.

In all cases a common mask was applied to the different datasets to avoid any issues of differences in spatial or temporal coverage. Since the GISTEMP has the least coverage of the datasets used here, all the other datasets were regridded to the GISTEMP resolution (2º by 2º) using a spline-interpolation and a common space-and-time mask was applied. In all cases the Arctic is defined as the region north of 70ºN and the NH as the region between 0ºN and 90ºN. We also computed the metrics using two alternative definitions of the Arctic as being the region north of 60ºN and north of 80ºN, but found no significant change in the results in either test (Figure S1, S2, S3, S4). This may be expected given that there is a very high correlation between the Arctic SAT anomalies defined as north of 60ºN and north of 70ºN e.g. r=0.92, p<0.01 between these two definitions in the Had4krig_v2 dataset.

## 3. Results

Figure 1 shows the AA as defined by the difference in temperature anomalies between the Arctic and the NH ($A_1$) smoothed using a 21-year running-mean to highlight the long-term variability. In the annual-mean there is a strong, multi-decadal variability seen in the two observational datasets and the ERA20C reanalysis with a peak in AA around 1940, a minimum around 1970, and a strong increase in AA in the periods from 1910 to 1940 and from the 1980s to the present. All the metrics except $A_2$ agree that the current annual-average AA is the strongest it has been since records began. This is most apparent when we take a shorter window over which to assess the metrics, which is a consequence of the very fast pace of change in the Arctic in the last 20 years (Figure 2). However, this is not the case when we look at the seasonal variation in the strength of AA. In the two observational datasets the early warm period, in the mid-1930s, had as strong or stronger winter-time AA. The winter generally has much stronger AA than the summer, and we also see much greater variability in the strength of AA in the winter (Figure 1C, 1E). This is largely due to the thermodynamic stabilizing effect of melting ice on the Arctic summer climate. Although in the two century-long reanalysis products, both reach maxima in the early 21$^{st}$ century. In the summer it is only the 20CRv2c which has a strong (and negative) AA at any time. In all the other datasets the magnitude of summer AA is always close to zero.

On the annual average we can see there is very good agreement between the observational records with only a small negative bias in the GISTEMP temperature with respect to the Had4krig record. This bias is strongest during the early-warming period when observations were sparse. The ERA20C reanalysis closely matches the observations from the 1940s through to the present day, but has a clear bias in the early 20$^{th}$ century. In contrast, the 20CRv2c reanalysis has very little correspondence with the observed temperature differences throughout the 20$^{th}$ century. These relationships between the datasets can be seen in the corresponding Taylor plot (Figure 1B). Note that since the modern-era reanalyses (ERAInt, MERRA2, CFSR, and JRA55) cover only a short period after applying the 21-year smoothing, they were not included in the Taylor-plot analysis. While both the reanalysis show similar variability to the observational datasets, only the ERA20C has a good correlation with the observations (r=0.64, p<0.01).

The large differences between the reanalysis products in the early warming period are likely due to the sparseness of the observations in the Arctic at this time. This can be seen from the difference in consistency between the representation of the early warming and recent warming periods in the different reanalysis products. The same variables are assimilated throughout the full period of the two century-long reanalyses, but the number and representativeness of the observations varies considerably in that time. In the modern warming period the high density of observations strongly-constrains all the different reanalysis products whether they include satellite observations or only surface observations. As a result all the reanalysis and observation products produce similar magnitudes and temporal variability in the degree of AA during the recent warming period.

The 20CRv2c reanalysis gives a very different result to the other datasets in winter with only a non-significant correlation to the GISSTEMP record, principally due to the presence of a period of strong increase in AA from the 1950s to the late 1970s which was not found in the GISSTEMP, or any other, dataset. The ERA20C shows a similarly close correspondence to the observational datasets in winter as in the annual average with a similar pattern of multi-decadal variability, although there is a generally weaker AA than was found in the observational datasets. However, in the summer there is very little correspondence between any two datasets; even the two observational datasets have a non-significant correlation. The 20CRv2c reanalysis may be expected to deviate from observations as the underlying model has known problems with reproducing the SAT in the Arctic due to a misspecification of sea ice in coastal regions (Compo et al., 2011).

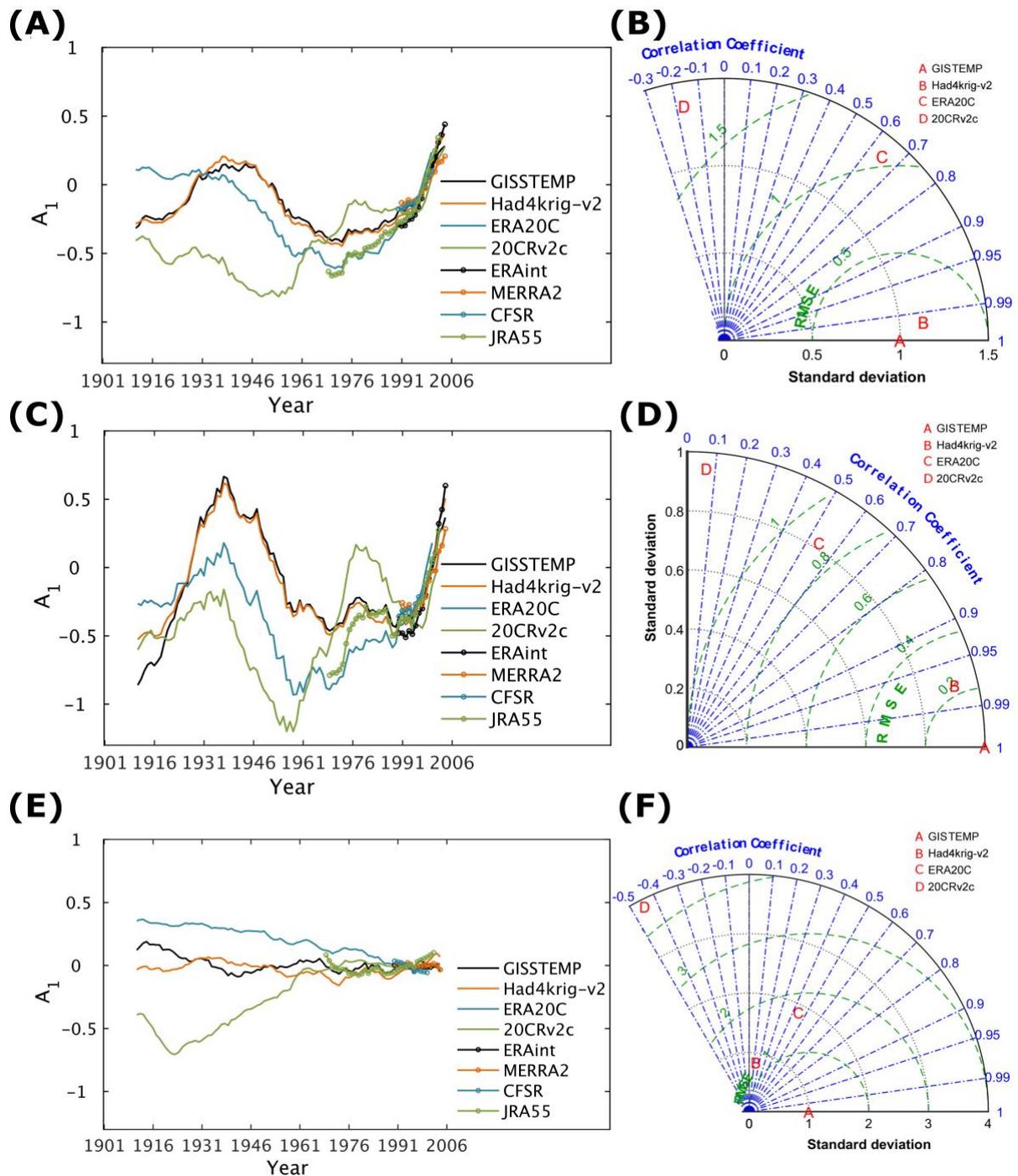

**Figure 1. (A)** The AA defined by the annual-average temperature difference between the Arctic temperature anomaly and the NH temperature anomaly smoothed using a 21-year running-mean ($A_1$). This is shown for the six reanalysis products: ERA20C; 20CRv2c; ERAint; MERRA2; CFSR; and JRA55, and the two observational datasets: GISTEMP and Had4krig_v2. **(B)** A Taylor plot of $A_1$ for the annual mean which shows the standard deviation of each of the time-series of the full-period datasets, and the Pearson correlation coefficient and root-mean-square of the errors (RMSE) between each series and the reference dataset, GISTEMP. The same analysis is repeated for the winter **(C)** and **(D),** and the summer **(E)** and **(F).**

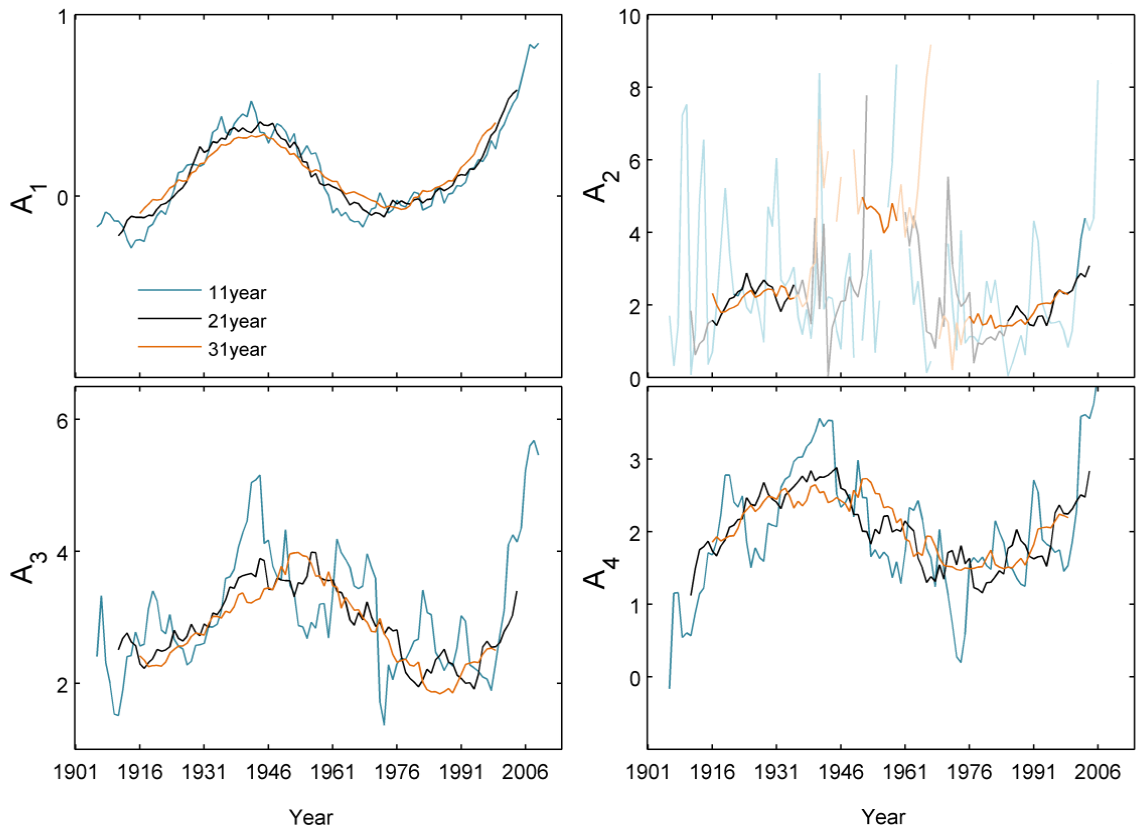

**Figure 2.** The four metrics for AA from the GISTEMP data calculated using three different lengths of running-window: 11, 21, and 31 years.

The ratio of the 21-year trends in the Arctic and NH, $A_2$, is shown in Figure 3. Due to the constraint that we require statistically significant trends in both the Arctic and NH in order to be able to obtain a value for this metric, there are large gaps in the time-series for all the datasets. We also removed all values greater than 10, as per Johannessen et al., (2016). There are two periods where there are relatively long records of AA that are statistically significant: from around 1915 to 1940 and from the mid-1980s to the present. In the most recent period we can see a general increase in the degree of AA from the mid-1980s to the present indicating an increasing rate of warming in the Arctic as compared to the hemispheric-average. While all the datasets indicate a strong AA in the mid-1950s, it is only the ERA20C dataset for which the values for the metric are statistically significant. At no point was there a statistically-significant value of $A_2$ less than or equal to one: so Arctic temperature trends were always found to be greater than the hemispheric average. During the 1990s the Arctic warming was approximately twice as strong as the hemispheric average and that has been increasing in recent years according to all reanalyses data (Figure 3A, 3C). The most recent values of $A_2$ indicate that the current rate of warming in the Arctic is around three times greater than the hemispheric average and still increasing, although notably similar $A_2$ AA rates were found in the 1920s.

As with $A_1$, the seasonal analysis shows us that it is in the winter when we have the strongest AA (Figure 3C). In three of the datasets, the two observational datasets and the 20CRv2c, the current AA shows the Arctic winters warming at over 6 times the rate of the hemispheric-average. In these three datasets the most recent values of $A_2$ are the highest that were seen in the full period.

In the annual-average the two observational datasets have the best agreement, with a high correlation and similar variability (Figure 3B). However, in contrast to the results from the $A_1$ metric, both ERA20C and 20CRv2C reanalysis have close correspondence to the observational datasets with a high correlation with the GISTEMP record (both with r=0.80, p<0.05), and a similar temporal variability. In the seasonal analysis there are much larger differences between all the datasets. In the winter (Figure 3D) it is the 20CRv2c which has the highest correlation to the GISSTEMP record (r=0.78, p<0.05), although the greater temporal variability means that it has a similar RMSE with respect to the GISSTEMP record as the other observational record, Had4krig-v2. In contrast the ERA20C shows similar variability to the GISSTEMP record, but a non-significant correlation. In the summer there is a generally weak AA with most values of $A_2$ lying between 0 and 2. While there is a good correlation between all the datasets (Figure 3F) there is a large difference in the magnitude of the variability, leading to poor overall consistency between the 20CRv2C and the other datasets.

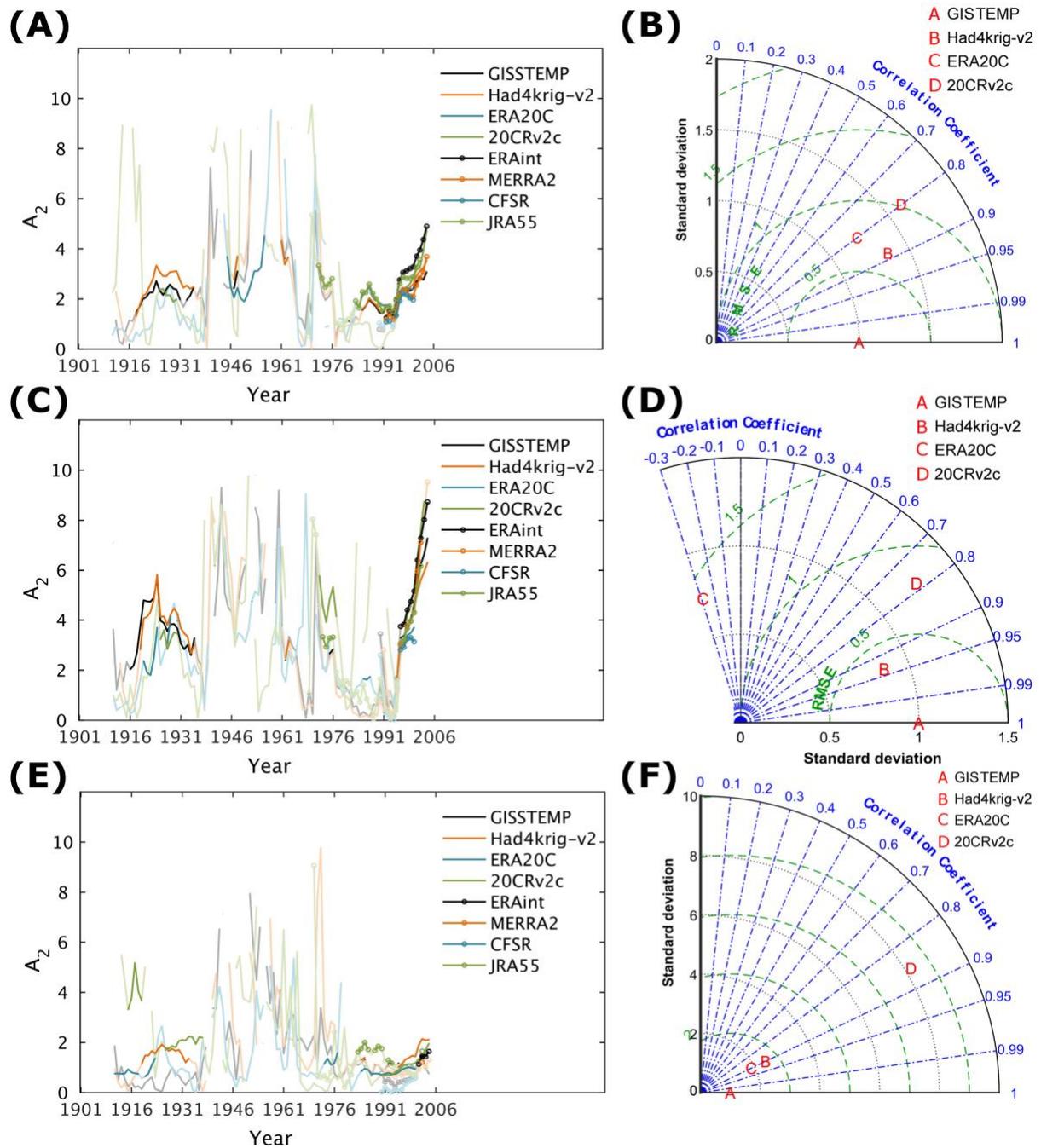

**Figure 3. A)** The measure of AA defined by the ratio of the absolute value in the 21-year linear trend in Arctic and NH SAT anomalies ($A_2$). This is shown for the six reanalysis products: ERA20C; 20CRv2c; ERAint; MERRA2; CFSR; JRA55, and the two observational datasets: GISTEMP and Had4krig_v2. Values greater than 10 were removed and those times when both trends are significant (p<0.05) are shown in solid colour with the non-significant values shown with lower opacity. **B)** A Taylor plot of the time series of $A_2$ which cover the full period with the GISTEMP series taken as the reference. The same analysis is repeated for the winter **(C)** and **(D),** and the summer **(E)** and **(F).**

The third metric, $A_3$, the ratio of inter-annual variability (the standard deviation of the anomalies in the SAT) in the Arctic and the NH is given in Figure 4. In the observational datasets there is a general increase in $A_3$ from the start of the 20$^{th}$ century until around 1960, followed by a decrease until the early 1990s until there is a sudden switch to a rapid increase in AA from the early 1990s continuing to the present. This pattern is seen in the both of the reanalyses, although the timing of the decrease in AA is less clear in the 20CRv2c reanalysis. In the minima, Arctic inter-annual variability is around twice as large as the hemispheric average and in periods of strong AA it may be three or four times larger. Compared to observational datasets, shorter-period reanalyses (e.g. ERAint and JRA55) tend to show a more rapid increase in AA in the 1990s, both in the annual-average and the wintertime (Figure 4A, 4C). The strongest AA was found in the winter with $A_3$ in the observational datasets peaking in the most recent years at values around 5 (Figure 4C). The observational datasets and the ERA20C also show a similar multi-decadal pattern in the winter as we see in the annual-mean, although the increase in AA in the early 1990s is much more rapid in winter. The highest summer-time AA in the observational datasets occurred in the 1960s (Figure 4E). The summertime AA shows a much smoother multi-decadal variability with the observational datasets having a smooth increase in $A_3$ from the start of the 20$^{th}$ century until the 1960s, a decline until the mid-1990s, followed by an increase into the 21$^{st}$ century. The ERA20C reanalysis has a similar temporal structure to the two observational datasets, but the 20CRv2c is very different with a general decrease in the summer AA throughout the whole period.

In the annual-average and wintertime the results are highly consistent between the two observational datasets, especially since the 1960s. This can be seen in the corresponding Taylor plots where the correlation between the two observational datasets is very high (r>0.9, p<0.01) and there is only a small difference in the variability (Figure 4B, 4D). Some small differences may be expected due to differences in the stations and processing techniques used to generate the two datasets. In the annual average both of the reanalysis show similarly good correlation with the observational datasets (r>0.63, p<0.01 and r=0.62, p<0.01 for the ERA20C and 20CRv2C respectively), although the ERA20C has a greater variability than either of the observational datasets. However, in the seasonal analysis we can see large differences between the two century-long reanalyses. In the winter, while the ERA20C shows a good correlation with the GISSTEMP record (r=0.72, p<0.01), the 20CRv2c has a very poor correspondence (r=0.30, p<0.01) with no decrease in AA between the 1960s and 1990s. In the summer the 20CRv2c has an even worse correspondence to the observations with a non-significant correlation to the GISSTEMP record, while the ERA20C has a relatively higher correlation to the GISSTEMP than the other observational record (r>0.44, p<0.01). So while the 20CRv2c may appear to be a good choice when assessing AA using $A_3$ in the annual-average, it does a very poor job in reproducing the observed seasonal differences in $A_3$.

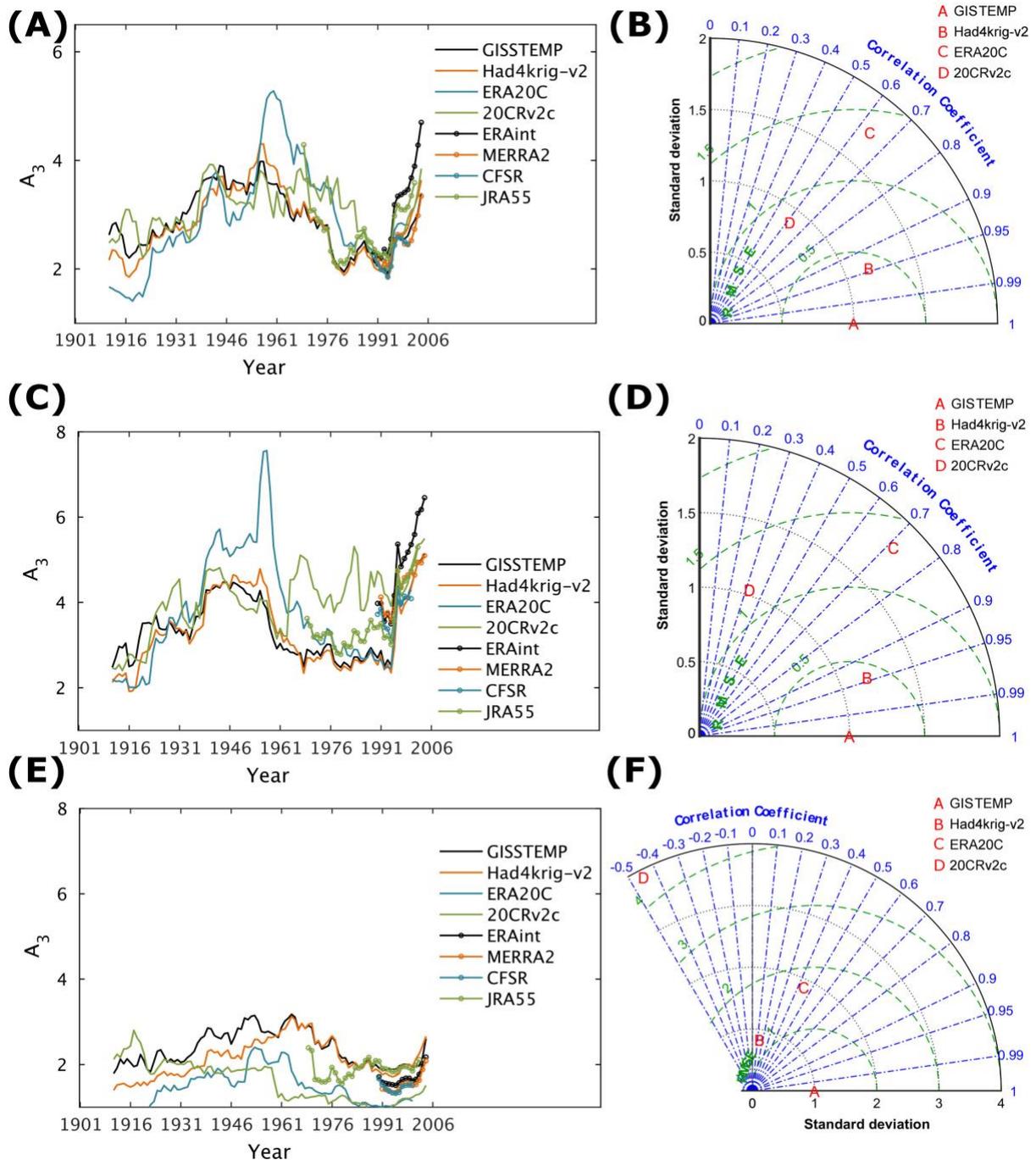

**Figure 4.** The AA defined by the ratio of the inter-annual temperature variability in a 21-year running-window between the Arctic and the NH ($A_3$). This is shown for the six reanalysis products: ERA20C; 20CRv2c; ERAint; MERRA2; CFSR; JRA55, and the two observational datasets: GISTEMP and Had4krig_v2. **(B)** A Taylor plot of $A_3$ for the full-period datasets using GISTEMP as the reference dataset. The same analysis is repeated for the winter **(C)** and **(D)**, and the summer **(E)** and **(F).**

The time-series of annual-averaged $A_4$ from the four datasets considered here are shown in Figure 5A. The two observational datasets show a high degree of similarity to the results from $A_1$ with a peak in the 1930s to 1940s, a minimum during the 1970s and increasing AA from around 1980 to the present day. Similar to $A_3$, ERAint tend to present more rapid increase in

AA during 1990s (as shown in Figure 5A, 5C). In the minima in the 1970s the value of $A_4$ is around 1, indicating that the magnitude of temperature anomalies is the same in the Arctic as in the hemispheric-average. We also note that $A_4$ shows the strongest magnitude of AA when we use the simultaneous SAT anomalies in the Arctic and in the NH i.e. there is no indication that either the Arctic or the NH is leading the AA (Figure S5). However, during periods of strong AA, Arctic anomalies are typically twice as strong as the hemispheric average with the most recent years having the strongest AA with Arctic anomalies currently around three times larger than the hemispheric average. There is a very good agreement between the observational datasets as to the temporal structure of $A_4$ ($r = 0.93$, $p<0.01$) and they have similar variability (Figure 5B). It is the 20CRv2c reanalysis which most closely matches the observations with $r=0.83$, $p<0.01$. While the ERA20C reanalysis has a very good match to the GISTEMP records on the temporal variability of $A_4$, the timing of the variability does not closely match that of GISTEMP ($r=0.47$, $p<0.01$). Despite the similarity between the results for $A_1$ and $A_4$ in the observational datasets, it should be noted that while ERA20C had a similar match to the observations in these two metrics, the 20CRv2c has a considerably better fit to the observations for $A_4$ than for $A_1$.

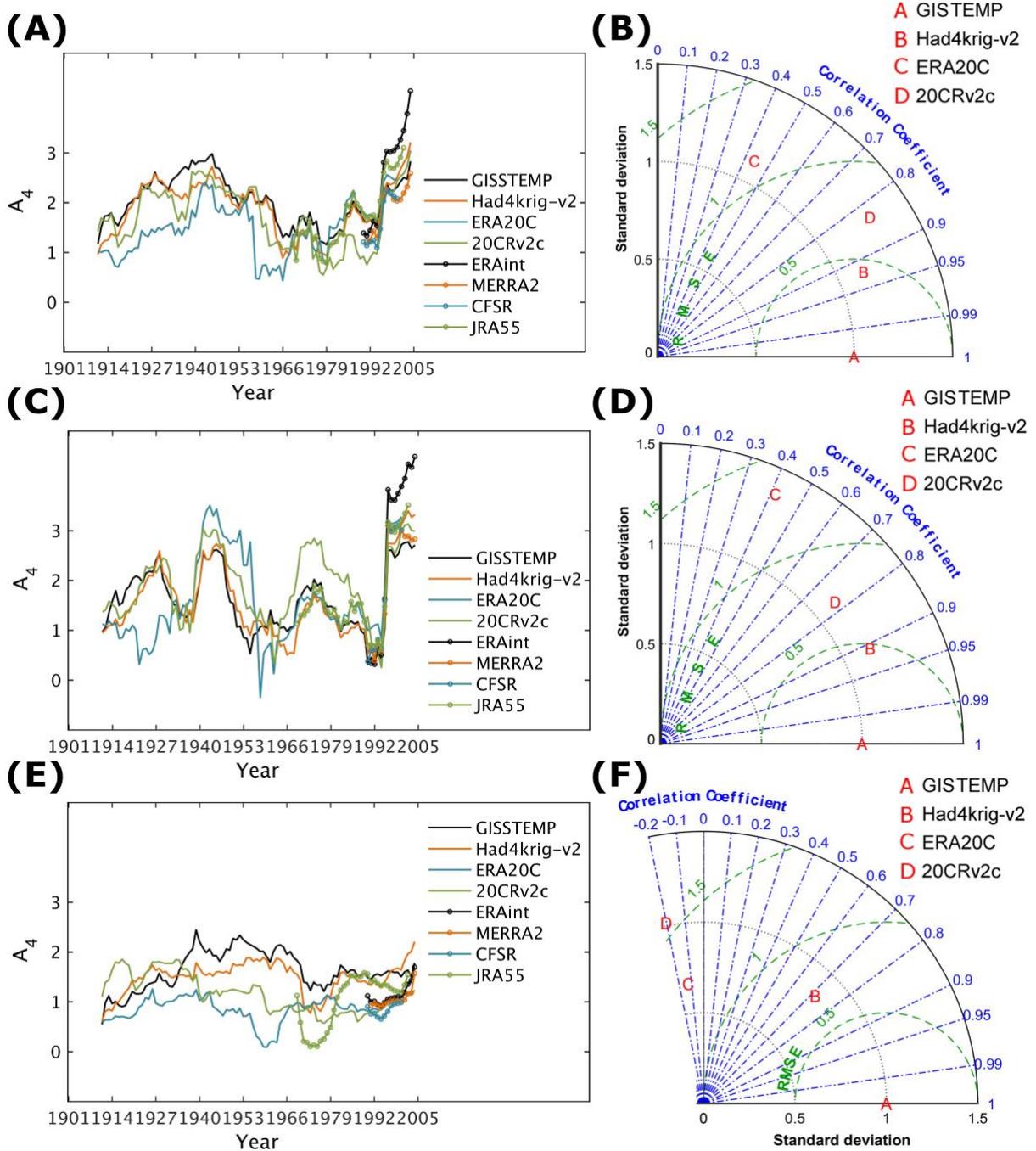

**Figure 5.** The AA defined by the ratio of the slope of the linear regression between Arctic and NH surface air temperature anomalies in a 21-year running-window ($A_4$). This is shown for the six reanalysis products: ERA20C; 20CRv2c; ERAint; MERRA2; CFSR; JRA55, and the two observational datasets: GISTEMP and Had4krig_v2. **(B)** A Taylor plot of $A_4$ for the full-period datasets using GISTEMP as the reference dataset. The same analysis is repeated for the winter **(C)** and **(D),** and the summer **(E)** and **(F).**

Of the two century-long reanalysis datasets, NOAA's 20CRv2c is closest to reproducing the results from the observations using the metrics based upon the SAT trends ($A_2$), variability ($A_3$) and regression ($A_4$) of the SAT anomalies. In contrast, the ERA20C reanalysis very closely matches the SAT anomalies ($A_1$) and variability of the SAT anomalies ($A_3$) for most of the 20$^{th}$ century. So, for example, if one wishes to assess the atmospheric dynamics during periods of 20$^{th}$ century AA using $A_2$ or $A_4$ then the 20CRv2c may be a better choice of reanalysis than the ERA20C, but both 20CRv2c and ERA20C would be good when using $A_3$.

All the metrics indicate that the period from around 1990 to the present is one of increasing AA, and this is also a consistent result across the different datasets. However, there are large differences in the interpretation of AA prior to the satellite era (1979-present) depending upon the choice of metric and dataset. In the observational datasets there was a peak in the AA around 1940 in the $A_1$ and $A_4$ metrics, whereas in $A_2$ and $A_3$ this period of AA didn't peak until the 1950s, although this cannot clearly be confirmed from $A_2$.

**4. Discussion and conclusion**

There are several aspects which should be considered when determining what metric to use to measure the degree of AA. First, it should be a metric which is especially sensitive to the process being studied while considering that, although the mean and standard deviation of surface air temperature are mathematically orthogonal, they are physically related (Esau et al., 2012). Second, one should avoid using ratios when the denominator is a metric which may approach zero, such as for trends or anomalies. Hind et al. (2016) have previously highlighted this problem, but we emphasize it here because it arose in our consideration of the ratio of trends, $A_2$. Third, it should be a metric of a variable which is well-characterised in the dataset being used. For example, the 20CRv2c reanalysis is reasonably consistent with the observations when it comes to the variability but has a very poor consistency with observations in the Arctic temperature anomalies. Fourth, one needs to consider the timescale that is relevant to the process being examined. The metrics related to the linear trends or inter-annual variability have non-independent values for each year due to the use of a running-window, and so they cannot be used to study temporal behaviour at periods shorter than the window length e.g. 21 years in the metrics presented here; whereas $A_4$ may be used to generate an independent value for each year or season and $A_1$ may be calculated at any temporal resolution. So if one wants to study the year-to-year variability in AA, then $A_1$ or $A_4$ may be the most appropriate choice of metric.

It is also important to consider the seasonal differences in the characterisation of AA in a given dataset. For example, the inter-annual variability in the 20CRv2c has a good match to the GISTEMP gridded-observations in the annual average, but this is due to a combination of a positive bias in the winter and a negative bias in the summer (e.g. for $A_3$ and $A_4$). So in a seasonal assessment of inter-annual variability it is the ERA20C which more closely matches the GISTEMP observations. However, in general there is less agreement between datasets as to the strength of AA in summer as AA is generally weaker in summer than in winter. The only exception to this was in the ratio of SAT trends for which there was a good agreement between all the datasets as to the temporal structure of the metric in summer, although there were large differences in the variability.

And finally, another important property of a metric of AA is that it equally compares climatic variations in the Arctic to those in the wider NH. For example, if we took the differences in the interannual variability in these two regions as our metric, this metric would be mainly determined by the changes in the Arctic due to the much higher variability in this region. So, although we have cautioned against the use of ratios when measuring AA, if the metric does not become sufficiently small to produce strong non-linearities, this can be an effective way to give equal weighting to the changes in the Arctic and the reference region. Another option would be to introduce an appropriate weighting, e.g. scaled to the magnitude of the variability in each region, so that the contribution to the value of the metric from each region is equal; such as is done for the standard normalised Azores minus Iceland NAO index.

**Acknowledgments**
The NASA Goddard Institute for Space Studies' "GISTEMP" data can be obtained from http://data.giss.nasa.gov/gistemp/. The European Centre for Medium Range Weather Forecasting "ERA20C" and "ERAint" data are publicly available and can be obtained from https://www.ecmwf.int/en/forecasts/datasets/browse-reanalysis-datasets. The "Had4Krig" dataset is the Hadley Cenre's HadCRUT4 dataset infilled by kriging and is publicly available from http://www-users.york.ac.uk/~kdc3/papers/coverage2013/series.html. The "20CRv2c" dataset is the 20$^{th}$ Century Reanalysis V2 data provided by the NOAA/OAR/ESRL PSD, Boulder, Colorado, USA, from their web site at http://www.esrl.noaa.gov/psd/. The Japanese 55-year reanalysis, JRA55, was produced and made publicly available by the Japanese Meteorological Agency. MERRA2 data were produced by NASA's Global Modeling and Assimilation Office and the data are hosted by http://disc.gsfc.nasa.gov/merra-2. The CFSR data were created by NOAA and are available from http://cfs.ncep.noaa.gov/cfsr/. The work was financially supported by the Research Council of Norway through the EuropeWeather project (no. 231322 / F20).